\documentclass[10pt,journal,compsoc]{IEEEtran}
\newif\ifpeerreview

\peerreviewfalse

\usepackage[nocompress]{cite}
\usepackage{url}
\usepackage{amsmath,amssymb,graphicx}
\usepackage[switch]{lineno}

\title{Learning Spatially Varying Pixel Exposures for Motion Deblurring}

 \author{Cindy~M.~Nguyen,		
		Julien~N.~P.~Martel,	and~Gordon~Wetzstein
\IEEEcompsocitemizethanks{\IEEEcompsocthanksitem C. M. Nguyen, J. N. P. Martel, and G. Wetzstein are with the Department of Electrical Engineering, Stanford University. \protect\\}
}

\begin{document}

\IEEEtitleabstractindextext{%
\begin{abstract}
Computationally removing the motion blur introduced by camera shake or object motion in a captured image remains a challenging task in computational photography. 
Deblurring methods are often limited by the fixed global exposure time of the image capture process. The post-processing algorithm either must deblur a longer exposure that contains relatively little noise or denoise a short exposure that intentionally removes the opportunity for blur at the cost of increased noise. 
We present a novel approach of leveraging spatially varying pixel exposures for motion deblurring using next-generation focal-plane sensor--processors along with an end-to-end design of these exposures and a machine learning--based motion-deblurring framework. We demonstrate in simulation and a physical prototype that learned spatially varying pixel exposures (L-SVPE) can successfully deblur scenes while recovering high frequency detail. Our work illustrates the promising role that focal-plane sensor--processors can play in the future of computational imaging.
\end{abstract}

}

\maketitle%

\IEEEraisesectionheading{
  \section{Introduction}\label{sec:introduction}
}
%
%
%
%
\IEEEPARstart{C}{ameras} have become ubiquitous, their presence felt in every smartphone and countless other devices sold today. Whether their images are designed for social media or processed by a self-driving car, modern cameras are still susceptible to the age-old problem of motion blur. This artifact is the result of either object or scene motion during the image exposure or camera shake via handheld photography, thus rendering the picture of a touching memory or an image used for computer vision tasks useless.

Unfortunately, removing motion blur remains an arduous task. Due to the heterogeneity of local and global motion blur, blind deblurring is difficult to address with pure deconvolution. Recent deep learning methods have popularized multi-scale~\cite{zamir2021multi,nah2017deep,suin2020spatially,tao2018scale} or multi-temporal~\cite{park2020multi,jin2018learning} deep learning approaches to address this issue. The coarse-to-fine nature of these networks allows for the gradual refining of motion deblurring kernels that are applied spatially invariantly, even in cases of non-uniform motion blur.

These methods have demonstrated the power of machine learning for deblurring, but notably do not take advantage of offloading computation to the hardware. On the other hand, computational imaging methods have excelled in combining software and hardware solutions to simplify many ill-posed computer vision tasks. These include engineering point spread functions (PSFs)~\cite{levin2008motion,elmalem2020motion,yosef2021video} and custom coded exposures~\cite{raskar2006coded,gu2010coded,tai2010coded,agrawal2009coded,jiang2021hdr} for motion deblurring. However, these methods are often limited to heuristic designs, and often restricted by fabrication limits, sensor capabilities, or human intuition. 

Fortunately, we are on the brink of a new era of sensor design. The rise of programmable sensors~\cite{carey2013100,zarandy2011focal,wong2020analognet}, in which sensing and processing can be executed together on the same silicon chip, has brought forth a new age of computing that operates at the time of image capture. Programmable sensors, or focal-plane sensor--processors, open up avenues for analog, digital, or mixed signal processing directly in-pixel. These sensors provide opportunities to preserve optical information of a scene that is otherwise irreversibly destroyed during the sensor integration or capture process. Moreover, the in-pixel intelligence running on these sensors can be learned using end-to-end (E2E) design strategies that jointly optimize the in-pixel programs with downstream computer vision algorithms.

\begin{figure}[t]
    \begin{center}
    \includegraphics[scale=0.7]{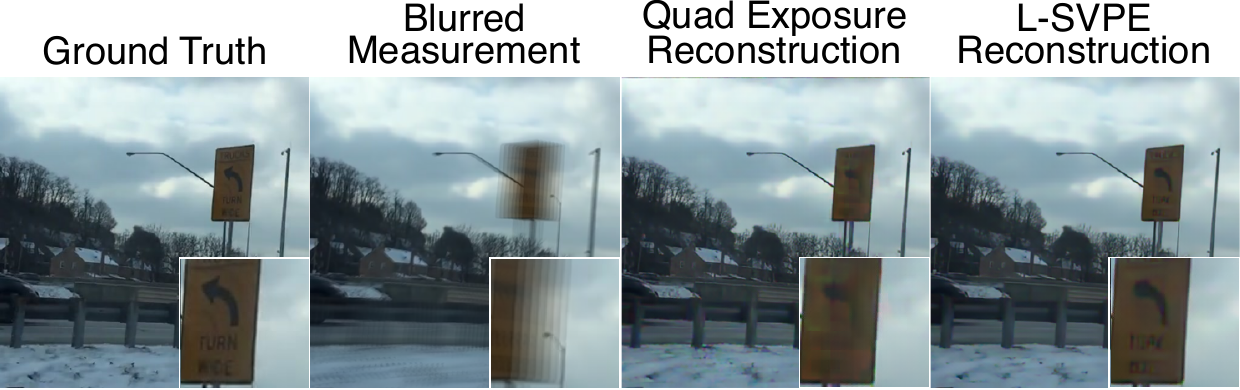}
    \end{center}
    \caption{An example reconstruction. The ground truth is the first frame of an input video segment, and the blurred image is all eight frames of the video segment averaged, which is the same as a 30 fps capture. The reconstruction from the \textit{Quad} exposure~\protect\cite{jiang2021hdr} with bilinear interpolation shows more motion and color artifacts compared to that of the learned spatially varying pixel exposure (\textit{L-SVPE}).}
    \label{fig:teaser}
\end{figure}

Given these capabilities, we address the above concerns with our jointly Learned Spatially Varying Pixel Exposures (L-SVPE) and machine learning--based reconstruction network to perform high quality motion deblurring (Fig. \ref{fig:teaser}). The benefits are two-fold. Within short exposures, the low signal-to-noise ratios are challenging to handle, but these short exposures still retain high frequency information that we desire in a reconstruction. Within longer exposures, images will have reduced noise levels thanks to signal accumulation. However, as the sensor integration time increases, these longer exposures are more prone to motion blur. We seek to use the complementary information from both in a single snapshot in our E2E framework.

As programmable sensors grow in production and popularity, we believe in the need to develop appropriate computational imaging algorithms to take advantage of these new capture capabilities. Future mobile phones may no longer be limited to the fixed global exposures they have today for capturing scenes with dynamic motion or low lighting. 
High dynamic range (HDR) imaging, most popularly done with some form of exposure bracketing~\cite{debevec2008recovering}, will struggle less with image alignment since programmable sensors can capture information at multiple exposures in a single snapshot~\cite{martel2020neural,so2021mantissacam}. We demonstrate that motion blur is just one of many tasks that highlight the utility of these powerful sensors.

Specifically, we make the following contributions:
\begin{itemize}
\item We introduce a fully differentiable model with a learned in-pixel encoder and deep convolutional decoder for motion deblurring. The encoder can be implemented on a programmable sensor offering ``in-pixel intelligence.''
\item We demonstrate in simulation that the E2E optimization of the coded exposures along with the decoding network yields superior reconstructions of motion blurred images over those of non-optimized exposures.
\item We implement a prototype camera using the SCAMP-5 programmable sensor--processor and demonstrate that our results translate to real-world captures, where the coded exposures are realized electronically.
\end{itemize}

Source code and trained models will be made available upon publication.

\section{Related Work}
\textbf{Motion deblurring.}
Early examples of blind motion deblurring include learning motion blur kernels using CNNs~\cite{sun2015learning,chakrabarti2016neural,schuler2015learning}. 
More recent examples increase the receptive field by using multi-resolution inputs~\cite{zamir2021multi,nah2017deep,suin2020spatially,tao2018scale} to learn on a combined global and local scale. 
Attention~\cite{suin2020spatially} and atrous convolutions~\cite{purohit2020region,miao2018aggregated} can help apply spatially varying weights for local blurs.
Inspired by transformers, Tu et al.~\cite{tu2022maxim} use multilayer perceptrons (MLPs) and attention with multi-resolution features for deblurring, reducing the required number of learnable parameters.
Though these methods have produced plausible reconstructions, deblurring networks are fundamentally limited due to the ill-posed nature of the problem, which can only be overcome by changing the image formation model through methods like E2E optimization.

\textbf{End-to-end optimization.}
Jointly designing optics and reconstruction networks has emerged as an incredibly useful paradigm in computational imaging~\cite{Wetzstein:2020}.
This E2E optimization has been applied to a number problems such as extended depth of field~\cite{sitzmann2018end}, depth estimation~\cite{chang2019deep,ikoma2021depth,wu2019phasecam3d}, HDR~\cite{metzler2020deep,sun2020learning}, super resolution localization microscopy~\cite{nehme2020deepstorm3d}, lensless imaging~\cite{sinha2017lensless,khan2020flatnet}, and Fourier ptychographic microscopy~\cite{kellman2019data}.
However, once the optics are fabricated, the challenge of calibrating them arises. 
To alleviate the need for calibration, in-pixel sensing strategies can also be designed in an E2E fashion for general capture~\cite{chakrabarti2016learning}, HDR~\cite{martel2020neural,so2021mantissacam}, and compressive imaging~\cite{vargas2021time}. We propose following this route, learning a spatially varying exposure pattern jointly with a network for the task of motion deblurring.

\textbf{Computational imaging for deblurring.}
Deblurring can be made less of an ill-posed problem with a variety of hardware modifications.
Raskar et al.~\cite{raskar2006coded} use a random binary coded global exposure pattern, implemented via a liquid-crystal shutter.
The coded exposure provides a PSF which preserves high spatial frequencies, allowing the blur to be decoded using traditional deconvolution algorithms.
Agrawal and Xu~\cite{agrawal2009coded} further improve this design with heuristics on PSF invertibility and estimation.
Other works use a custom rolling shutter~\cite{gu2010coded} or deblur using information from a camera's default rolling shutter~\cite{su2015rolling}.
Jeon et al.~\cite{jeon2017multi} use communication theory to design fluttering patterns, while more recently, Jiang et al.~\cite{jiang2021hdr} use an interlaced short, medium, and long exposure pattern with a specialized network for reconstruction. We compare this exposure pattern with ours, while using a simpler network for reconstruction.
Elmalem et al.~\cite{elmalem2020motion} design an optic that encodes blur in its color PSF, and Yosef et al.~\cite{yosef2021video} build on this work by additionally using a focus mechanism to recover video frames from a capture with motion blur.
Rengarajan et al.~\cite{rengarajan2020photosequencing} proposed using short-long-short exposures with recursive blur decomposition to deblur.
Notably, all the above approaches have been designed using heuristics and theory, making the search space limited to human intuition.


\textbf{Programmable sensors.}
The emergence of programmable sensors, offering unprecedented flexibility of in-pixel processing, has brought forth numerous interesting ideas for computational photography.
These sensors, also known as focal-plane sensor--processors~\cite{zarandy2011focal}, conduct low-level image processing during the capture. They reduce the need for excessive computational post-processing and enable new capture processes of an image to preserve information that would be lost otherwise, such as dynamic range. 
Programmable sensors, such as the SCAMP-5~\cite{carey2013100} which we use as our prototype, offer programmable pixels and have been successfully used in HDR imaging~\cite{martel2020neural}, video compressive imaging~\cite{martel2020neural,iliadis2016deepbinarymask}, depth from defocus~\cite{martel2017high}, feature classification~\cite{chen2017feature}, and ego-motion estimation~\cite{bose2017visual}.
More recently, coded exposures have also been used for compressive light-field and hyperspectral imaging~\cite{vargas2021time}.
Here, we propose to use one variant of these new vision chips, the SCAMP-5, to program pixel-wise coded exposures.
To our knowledge, this is the first application of these sensor--processors to motion deblurring.

\section{Formulation}
\begin{figure*}[t!]
    \begin{center}
    \includegraphics[width=\textwidth, scale=0.5]{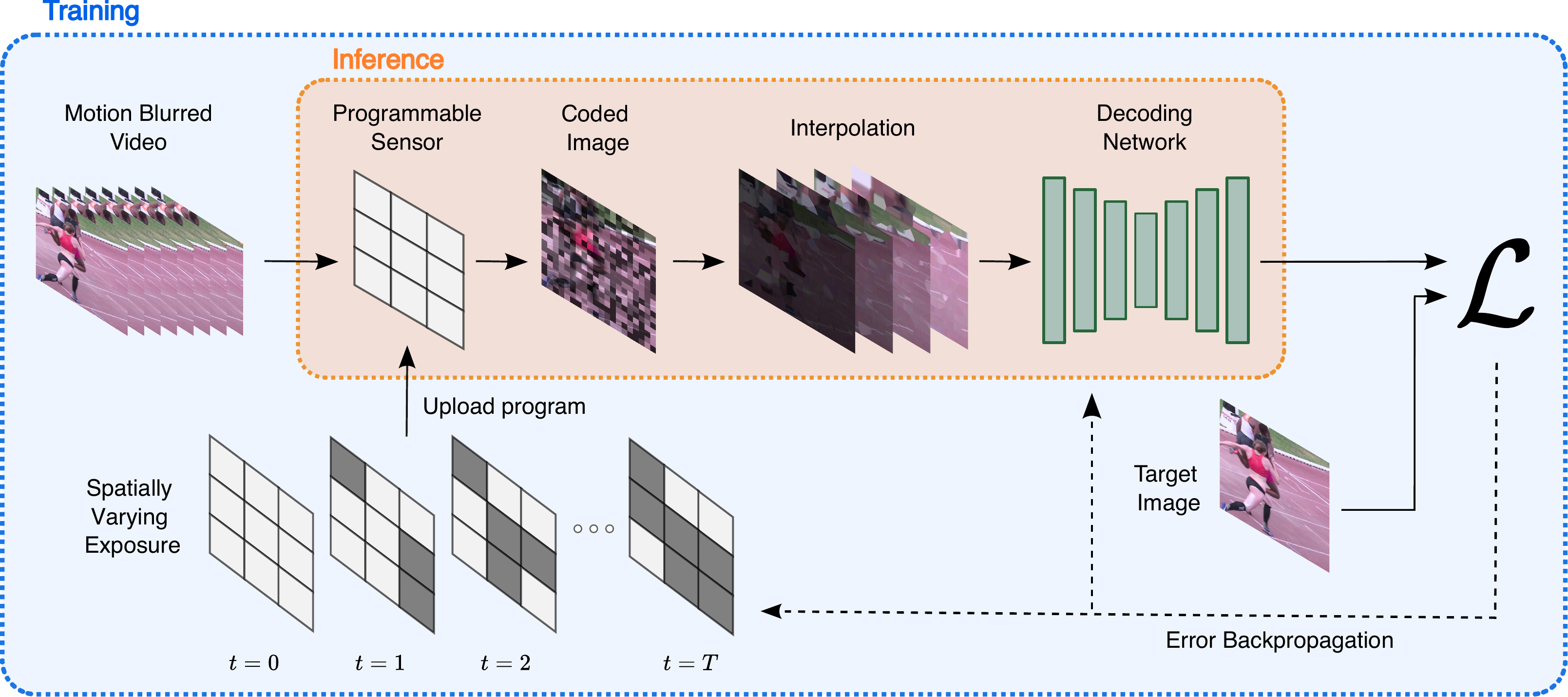}
    \end{center}
       \caption{Illustration of our \textit{L-SVPE} optimization framework, including the learned coded shutter, interpolation step, and decoding network. The error between the reconstructed image and ground truth is backpropagated to the coded exposure lengths and decoding network. Here, $T$ represents the maximum exposure length. This pipeline illustrates an example use of four different exposure lengths, providing four full-resolution interpolations.}
    \label{fig:pipeline}
\end{figure*}

 We simulate the capture of a scene with our learned programmable sensor using spatially varying pixel exposures (Sec. \ref{method:encoder}), which we also refer to as coded exposures. We then form a multi-channel image with $C$ channels from the single snapshot, where $C$ represents the number of unique exposure lengths in the learned coded exposure. We do so by interpolating (Sec. \ref{method:interp}) exposure pixels that were not explicitly captured. This image stack with a resolution of $H \times W \times C$ pixels is fed as the input of the decoding network (Sec. \ref{method:decoder}) to produce a reconstructed image. The pipeline is illustrated in Fig. \ref{fig:pipeline}.

\subsection{Learned spatially varying pixel exposures}
\label{method:encoder}
We model the exposure at pixel location $p$ as the integration of the incident irradiance $V_p$ over an exposure time $\Delta t$, which can be written as 

\begin{equation}
    \label{eqn:rad}
    E_p(t) = \int_t^{t + \Delta t} V_p(t') dt'.
\end{equation}
Here, we let $E_p$ represent the exposure at pixel $p$. We introduce a learned spatially varying exposure time $\Delta p$ for each pixel $p$ to indicate an ``on'' (contrary to ``off'') shutter.
Thus Eq. \ref{eqn:rad} can be rewritten as

\begin{equation}
    \label{eqn:shutter}
    E_p(t) = \int_{t}^{t + \Delta p} V_p(t') dt'.
\end{equation}
The exposure $E_p$ relates to the captured image via the camera response function as $I_p(t) = \mathcal{R}(E_p(t))$, which captures the noise and quantization effects of the camera.

\subsection{Exposure length--specific interpolation}
\label{method:interp}
The single-shot measurement on our focal-plane sensor--processor is a single-channel, grayscale image. We expand the measurement as a preconditioning step before decoding. Since our learned exposure times will vary spatially, we interpolate the missing exposure pixels to generate a full resolution estimate of the utilized exposure lengths. 
In our framework, we discretize the continuous varying exposure time $t$ to a set of discrete values $\mathcal{S} = \{1, .., T\}$, in which $T$ is the maximum exposure time.
We then interpolate all pixels of the single-channel sensor image with the same exposure to form a separate channel of the image stack that serves as the input to our motion deblurring network. 
We argue this interpolation step speeds up training since the learned kernels in the decoding network can be then applied spatially invariantly. 
We apply the interpolation function $\mathcal{U} : \mathbb{R}^{H \times W} \rightarrow \mathbb{R}^{H \times W \times C}$ where $C$ is the unique number of exposure lengths used from $\mathcal{S}$.

We use two different types of interpolation: Bilinear and Scatter-weighted interpolation. Bilinear interpolation (denoted as \textbf{B}) is applied in cases where the exposures have a $2 \times 2$ (Quad) or $3 \times 3$ (Nonad) tiled arrangement (see Sec. \ref{sec:baselines}), using information from regular sampling grids. 
Scatter-weighted interpolation (denoted as \textbf{S}) is applied in cases where pixel exposures of the same length have varying distances from each other across the sensor.
Scatter interpolation is applicable when the coded exposures are random or learned, but this method can also be used in the tiled case. 
Scatter interpolation requires finding the $N$ closest neighbors using a fast \textit{k}-d tree~\cite{kanungo2002efficient} and adding the values of neighboring pixels together, each weighted by their inverse distance to the point of interest to some power $r$. 
The interpolated value at point $p$ can be written as

\begin{equation}
\mathcal{U}(I_p) = \frac{\sum_{i=1}^N w_i(p) I_i}{ \sum_{i=1}^N w_i(p)},
\end{equation}
where $I_i$ is the value at neighboring pixel $i$. The weighting function $w_i(p) = \frac{1}{d(p, p_i)^r}$ is computed using Euclidean distance as $d$. Also, $N$ represents the number of nearest neighbors sharing the same exposure length as pixel $p$.

\subsection{Capture decoding}
\label{method:decoder}
Once we obtain an interpolated multi-channel image, we decode this using the well-known U-Net architecture~\cite{ronneberger2015u}. Our CNN $W_\psi : \mathbb{R}^{H \times W \times C} \rightarrow \mathbb{R}^{H \times W}$ contains skip connections with a depth of 6 without batch normalization, and each downsampling block contains a single convolution and ReLU with learned parameters $\psi$. The upsampling blocks use ConvTranspose to upsample the features at each stage. Our decoding step can be written as 

\begin{equation}
\hat{Y_p} = W_\psi (\mathcal{U}(I_p)),
\end{equation}
where $\hat{Y_p}$ is the reconstructed grayscale value at pixel $p$. We opt for the widely used U-Net architecture that has been useful for a variety of imaging problems~\cite{cho2021rethinking,lim2017enhanced} and has less parameters than the state-of-the-art deep deblurring methods~\cite{zamir2021multi}.

\subsection{Loss}
\label{method:loss}
We use a linear combination of an MSE loss ($L_2$) and a perceptual loss using VGG features ($L_{\textrm{VGG}}$)~\cite{zhang2018unreasonable}. Our loss can be written as

\begin{equation}
    L_{\text{percep}} = L_2 + \lambda L_{\textrm{VGG}},
\end{equation}
where $\lambda$ is the coefficient to weight the VGG-based loss. We use $\lambda = 100$. 

We compute the loss between the reconstruction with the clean first frame of the video segment, which we use as ground truth. 
We choose the first frame to reduce the need to deblur and prioritize using the information from the shortest exposure, as denoising is typically easier than deblurring.

\begin{figure}[t!]
    \begin{center}
    \includegraphics[scale=0.8]{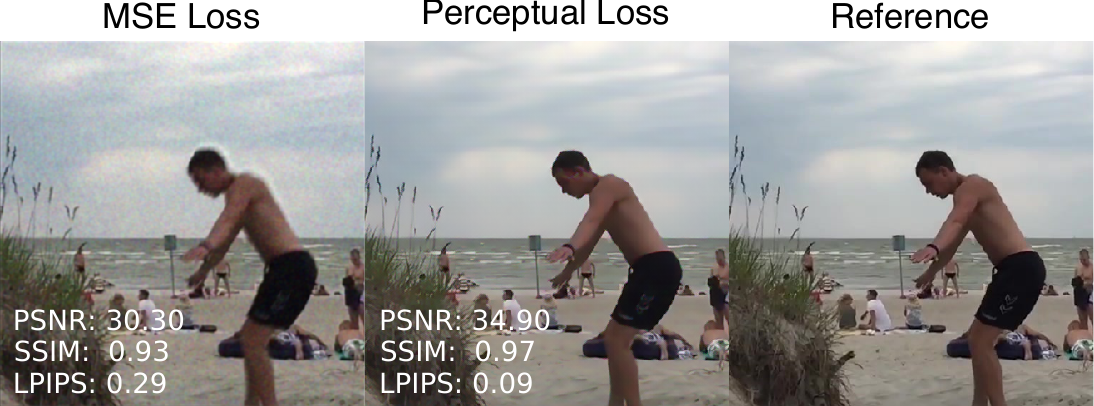}
    \end{center}
       \caption{Qualitative comparison between the reconstructions of two learned exposure models using \textit{L-SVPE} trained with an $L_2$ loss and our perceptually-guided loss. As LPIPS decreases, the quality of the reconstruction increases. $L_2$ lends itself to more blurry reconstructions, which may miss high frequency details.}
    \label{fig:loss}
\end{figure}

The VGG-based loss is exceedingly helpful in reaching the perception--distortion trade-off~\cite{blau2018perception,ledig2017photo}, which describes networks as trading off PSNR and SSIM performance for perceptual quality. Since there are many plausible reconstructions for motion blurred images, we find that the perceptual loss helps find a more suitable reconstruction over an unregularized MSE loss, which optimizes for per-pixel accuracy (Fig. \ref{fig:loss}).

\section{Implementation}
\textbf{Dataset.} To train our network and evaluate its performance, we use the Need for Speed (NfS) dataset~\cite{kiani2017need}, which consists of 100 videos obtained from the internet. 
Each video is captured at 240 frames per second (fps) with a $1280 \times 720$ resolution that we center crop to $512 \times 512$. 
We allocate 80 videos for our training set and 20 videos for our test set. 
For each video, we select 8 random 8-frame-long segments within the video, and each segment is averaged per-pixel to simulate a 30 fps capture.
This processing produces 640 video segments for training and 160 video segments for testing. 
We augment these segments during training by introducing random flips and rotations, and normalize the images to be within $[0, 1]$. 
Our input to the E2E model is an 8-frame video segment, and we use the first frame of the video segment, equivalent to a Short exposure, as the ground truth frame. 

\textbf{Training.} Our model was implemented in PyTorch and trained using an NVIDIA Quadro RTX 6000 GPU with 24 GB.
Our model was trained for 1000 epochs with the AdamW ~\cite{loshchilov2017decoupled} optimizer ($\beta_1=0.9, \beta_2=0.999$). For networks with learned coded exposures, we use an exposure learning rate of $2 \times 10^{-4}$ and a decoder learning rate of $5 \times 10^{-4}$. For baselines with fixed exposures, we use a decoder learning rate of $2 \times 10^{-4}$.

\textbf{Metrics.} We use three metrics to quantify the quality of our results. The first two are PSNR and SSIM~\cite{wang2004image} as traditional image quality metrics.
We also use the perceptual metric LPIPS~\cite{zhang2018unreasonable}.
We provide both quantitative and qualitative evaluation of our method compared to baselines (Sec. \ref{sec:baselines}), and we demonstrate the utility of each component in our ablation studies (Sec. \ref{sec:ablations}).

\section{Experiments}
We perform a series of experiments to highlight the value of our learned coded exposure. 
Specifically, we evaluate our method on different coded exposures and ablate several design choices made in our E2E model.

\begin{figure}[t]
    \begin{center}
    \includegraphics[scale=1.5]{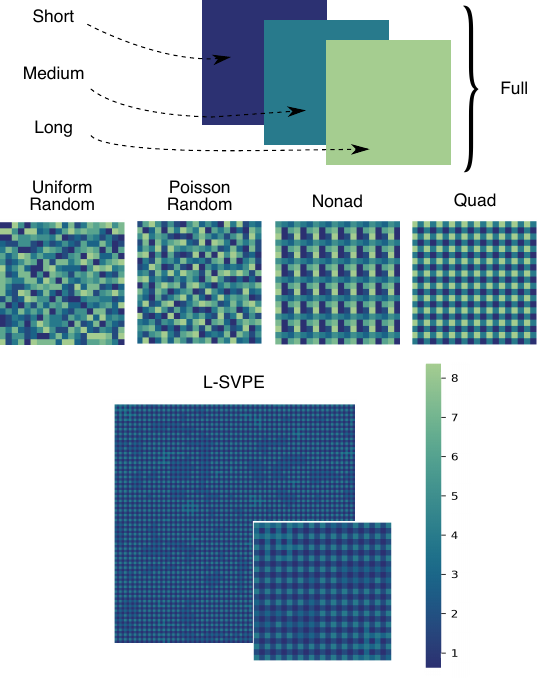}
    \end{center}
       \caption{Visualization of coded exposures of each baseline. \textit{Short}, \textit{Medium}, and \textit{Long} are fixed global exposures. The \textit{Full} consists of all three global exposures concatenated together. Each coded exposure (\textit{Uniform Random}, \textit{Poisson Random}, \textit{Nonad}, and \textit{Quad}) contains individual pixel exposures that can span from lengths 1 through 8, with cropped versions of the full resolution exposure shown here. We display the full resolution and a crop of the \textit{L-SVPE} exposure to highlight the differences from \textit{Quad}. Note that the average pixel exposure of \textit{L-SVPE} is lower than that of \textit{Quad}.}
    \label{fig:learned_shutters}
\end{figure}

\subsection{Baselines}
To demonstrate the utility of our learned exposure, we compare its performance against the following fixed exposures, where \textbf{B} represents Bilinear interpolation and \textbf{S} represents Scatter interpolation. Our baselines (Fig. \ref{fig:learned_shutters}) include:

\begin{itemize}
    \item \textbf{Burst Average}: All 8 frames are averaged together, simulating a Long exposure, and compared to the ground truth. This baseline does not use a decoder.
    \item \textbf{Short}: We simulate a 240 fps capture which is equivalent to a single frame of the video segment input.
    \item \textbf{Medium}: We simulate a 120 fps capture by averaging the first four frames of the input. 
    \item \textbf{Long}: We simulate a 30 fps capture by averaging all eight frames of the input.
    \item \textbf{Uniform Random (S)}: We initialize a fixed $512 \times 512$ array of pixel exposures uniformly selected from length 1 through 8.
    \item \textbf{Poisson Random (S)}: We use a multi-class Poisson disk sampling algorithm~\cite{wei2010multi} to generate a Poisson-distributed fixed pixel exposure from length 1 through 8.
    \item \textbf{Nonad (nine-tuple) (B, S)}: We use the full range of exposures in a $3 \times 3$ arrangement. We randomly arrange an array of pixel exposures, each pixel a unique length from 1 through 8, with an additional exposure length 1 to make a total of 9 pixels in the $3 \times 3$. We then tile this fixed arrangement to fill the $512 \times 512$ resolution.
    \item \textbf{Quad (B, S)}: Following Jiang et al.~\cite{jiang2021hdr}, we use a fixed tile coded arrangement of LMMS (long-medium-medium-short), where the short pixel is 240 fps, medium pixels are 120 fps, and long pixel is 30 fps. We use this baseline to also initialize L-SVPE which explains its resemblance (Fig. \ref{fig:learned_shutters}).
    \item \textbf{Full}: We concatenated a full-resolution stack of the Short, Medium, and Long exposure all captured with the same start time and same viewpoint. This baseline serves as theoretical upper bound for how well our method can do without requiring an interpolation step and with full resolution information at different exposure times. This stack of captures would be physically impossible to capture but can be easily simulated.
\end{itemize}

For fair comparison, we use the same decoder network as our method on all baselines except the Burst Average, which does not use a decoding network. L-SVPE, by default, uses Scatter interpolation. See the supplemental material for more details on our choice of baselines.

\begin{figure*}[t!]
    \begin{center}
    \includegraphics[width=\textwidth]{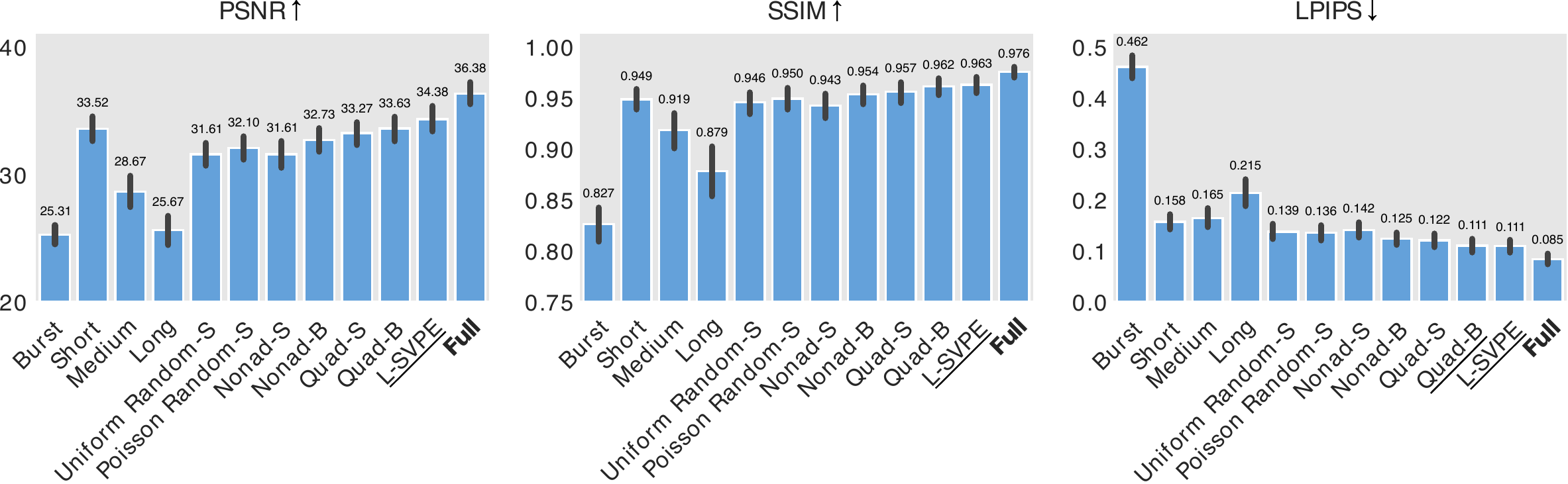}
    \end{center}
       \caption{Quantitative comparison of baselines. Printed values denote the numerical average, while gray bars represent the standard deviation. The \textit{Full} exposure represents a theoretical upper bound to our method, and \textit{L-SVPE} outperforms better than any of the fixed baselines, most closely reaching the theoretical upper bound on all metrics. These results are computed using the NfS dataset described in the text.}
    \label{fig:baselines_bar}
\end{figure*}

\subsection{Determining interpolation parameters}
To determine the optimal parameters for Scatter interpolation, we compute the time it takes to perform interpolation with $k$ neighbors. We compute the average time and accuracy of interpolating a test set of video segments captured with the chosen exposure. We first simulate the exposure capture of each video segment using the spatially varying exposures under the Exposure column to get a single-channel capture. We then use Bilinear or Scatter interpolation to interpolate pixels of exposure lengths that were not explicitly captured to acquire a $H \times W \times C$ multi-channel image. The Time column shows the time required in seconds to compute this step. We then compute the accuracy of each interpolation of each individual channel and average the metrics for each channel together for each image. We then compute the average of the metrics over the entire test dataset.

We present these results in Table \ref{table:comp_interp}. We test interpolation on the Quad and Nonad exposures. Bilinear interpolation (denoted as B) does not have a $k$ parameter. We observe that as $k$ increases, so does the time needed for computation per image. Scatter interpolation (denoted as S) is slower than Bilinear, but in the case of Quad, can provide more accurate interpolations. With Nonad exposure, we observe that Bilinear performs best, while Scatter degrades the quality of the reconstruction since neighbors are much further in the Nonad case. Thus, for all our Scatter-based methods, we use $N = 3$ and $r = 1$ which allows for relatively fast computation and accuracy.

\setlength{\tabcolsep}{2pt}
\begin{table}
\begin{center}
\caption{Comparison of interpolation methods. We show the efficiency and accuracy of Bilinear (denoted with B) and Scatter (denoted with S) interpolation methods with varying parameters. Time denotes the time required to interpolate a single image on average. Bilinear interpolation is the fastest for both \textit{Quad} and \textit{Nonad}. Scatter interpolation takes noticeably longer with the need to search for neighbors and is less accurate due to information dilution from further pixels. All Scatter-based methods here use $r = 1$.}
\label{table:comp_interp}
\begin{tabular}{lc|c|cccc}
\hline\noalign{\smallskip}
Exposure & $k$ Neighbors & Time (s) & PSNR $\uparrow$ & SSIM $\uparrow$ & LPIPS $\downarrow$ \\
\noalign{\smallskip}
\hline
\noalign{\smallskip}
Quad~\cite{jiang2021hdr} (B) & --- & \textbf{0.001} & 37.257 & 0.975 & \textbf{0.049} \\
Quad (S) & 3 & \underline{0.041} & \underline{42.138} & \textbf{0.988} & \underline{0.062}\\
Quad (S) & 4 &  0.042 & \textbf{42.826} & \textbf{0.988} & 0.063\\
Quad (S) & 5 & 0.044 & 41.914 & \underline{0.986} & 0.076\\
Nonad (B) & --- &  \textbf{0.001} & 30.733 & 0.837 & 0.089\\
Nonad (S) & 3 & 0.117 & 28.912 & 0.826 & 0.127\\
Nonad (S) & 4 & 0.124 & 28.783 & 0.825 & 0.150\\
Nonad (S) & 5 & 0.131 & 28.520 & 0.822 & 0.154\\
\hline
\end{tabular}
\end{center}
\end{table}
\setlength{\tabcolsep}{1.4pt}

\subsection{Comparison against baselines}
\label{sec:baselines}
Figure \ref{fig:baselines_bar} presents qualitative comparisons between these baselines. The Short exposure performs better than Medium and Long exposures, due to the network more easily learning to denoise than deblurring. The Full exposure outperforms all single exposure baselines, demonstrating the utility of combining information from different exposure lengths.

We see that baselines with spatially variant pixels outperform the Medium and Long baselines, while also notably performing better than the Short exposure signficantly in LPIPS. This result demonstrates the utility of varying exposures for perceptual quality. More structured shutters, like Nonad and Quad, do better than random shutters likely because of the uniformity of data given from the sensor. Our approach with learned exposures can outperform all baselines, reaching the closest to the theoretical upper bound, Full, in performance.

\subsection{Ablation studies}
\label{sec:ablations}
Table \ref{table:ablation} presents ablation studies on each component of the network. We compare our decoder choice, U-Net, against DnCNN~\cite{zhang2017beyond}, another popular memory-efficient image reconstruction network. The DnCNN is trained to predict the residual noise of the first channel of the multi-channel input into the decoder, if applicable.
These studies demonstrate the utility of each component of our method. Note that the Scatter interpolation may not provide the best PSNR performance over no interpolation, but improves SSIM and LPIPS, which is consistent with our baselines study results. 
We found that DnCNN with a perceptual loss can be often detrimental to performance, which may suggest that DnCNN is not universally optimal for providing perceptually plausible deblurring from spatially varying exposures.

\setlength{\tabcolsep}{2pt}
\begin{table}
\begin{center}
\caption{Ablation studies showing the effect of the choice of shutter, interpolation, network, and loss on performance. We use (S) to denote methods which use Scatter interpolation. No (S) denotes that no interpolation step was used.}
\label{table:ablation}
\begin{tabular}{lcc|ccc}
\hline\noalign{\smallskip}
Exposure  & Network & Loss & PSNR$\uparrow$ & SSIM$\uparrow$ & LPIPS$\downarrow$\\
\noalign{\smallskip}
\hline
\noalign{\smallskip}
Uniform Random & DnCNN~\cite{zhang2017beyond} & $L_2$ &  30.345 & 0.913 & 0.338\\
Uniform Random & DnCNN & Percep. & 12.269 & 0.272 & 0.763\\
Uniform Random (S) & DnCNN & $L_2$ & 30.325 & 0.920 & 0.319\\
Uniform Random (S) & DnCNN & Percep. & 28.255 & 0.915 & 0.225\\
Uniform Random & U-Net~\cite{ronneberger2015u} & $L_2$ & 33.149 & 0.950 & 0.181\\
Uniform Random & U-Net & Percep. & 28.877 & 0.905 & 0.200\\
Quad~\cite{jiang2021hdr} & DnCNN & $L_2$ & 31.435 & 0.918 & 0.355\\
Quad & DnCNN & Percep. & 11.719 & 0.247 & 0.752\\
Quad & U-Net & Percep. & 32.345 & 0.952 & 0.140\\
L-SVPE & DnCNN & $L_2$ & 30.025 & 0.860 & 0.369\\
L-SVPE & DnCNN & Percep. & 12.434 & 0.268 & 0.751\\
L-SVPE (S) & DnCNN & Percep. & 29.530 & 0.919 & 0.230\\
L-SVPE & U-Net & $L_2$ & 27.061 & 0.865 & 0.406\\
L-SVPE & U-Net & Percep. & \textbf{34.625} & \underline{0.961} & \underline{0.122}\\
L-SVPE (S) & U-Net & $L_2$ & 28.987 & 0.892 & 0.320\\
L-SVPE (S) & U-Net & Percep. & \underline{34.522} & \textbf{0.967} & \textbf{0.105}\\
\hline
\end{tabular}

\end{center}
\end{table}
\setlength{\tabcolsep}{1.4pt}

\subsection{Image reconstruction quality}
\begin{figure*}[t!]
    \begin{center}
    \includegraphics[width=\textwidth]{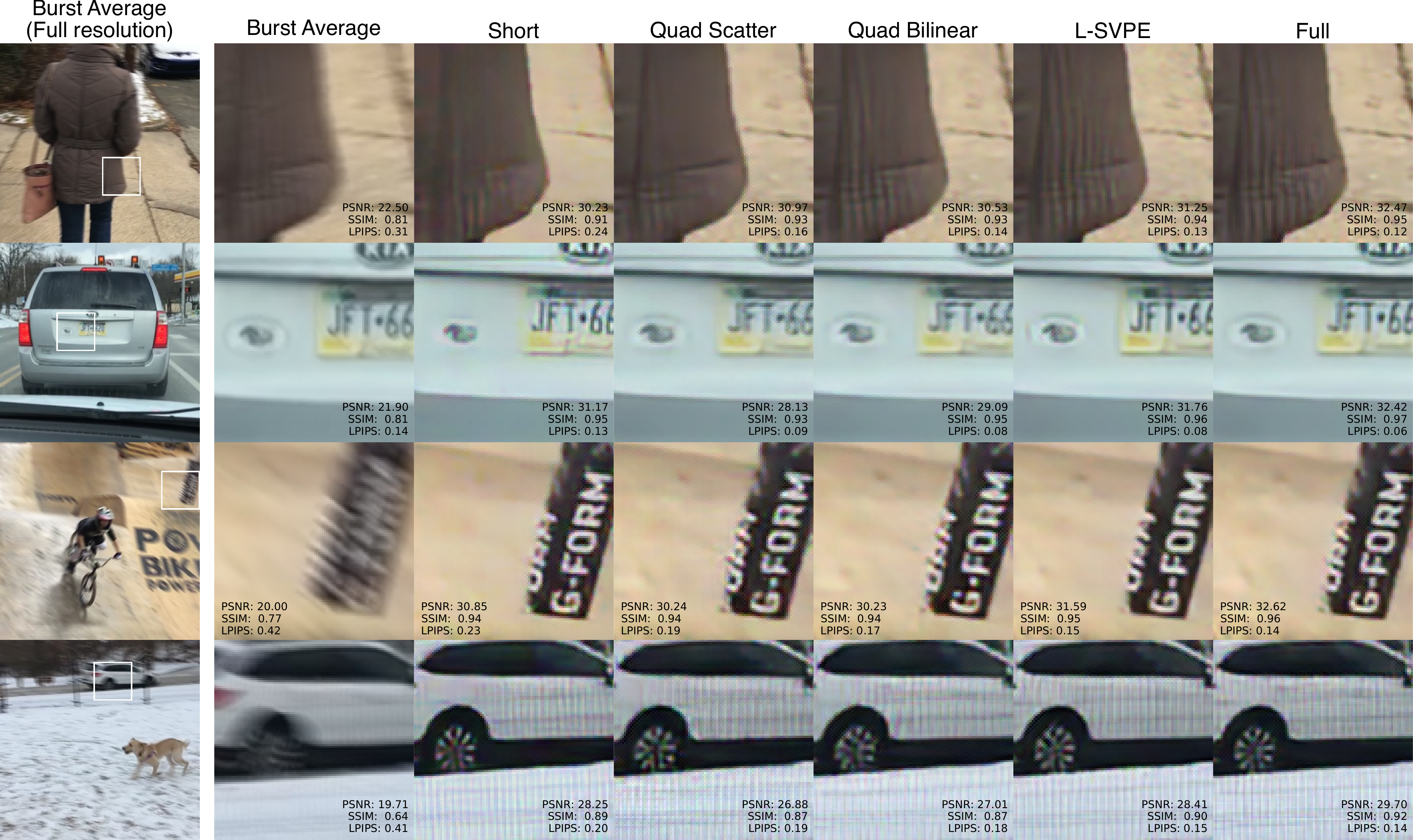}
    \end{center}
       \caption{Qualitative comparison of the five top-performing baselines. Each row shows an example frame of the full resolution average of all the video segment frames (\textit{Burst Average}), a $100 \times 100$ crop of a region of the \textit{Burst Average}, and the same crop of the reconstructions from the top five performing baselines. Reconstructions of all baselines are available in the supplemental material. 
       \textbf{Row 1:} The high frequency detail of the jacket is well recovered by \textit{L-SVPE} while blurred out by other methods. 
       \textbf{Row 2:} The \textit{Short} baseline fails to capture the border of the bumper sticker while introducing colored artifacts. \textit{L-SVPE} correctly recovers the border and provides sufficient clarity to the lettering of the license plate. 
       \textbf{Row 3:} The coloring artifacts can be observed in the \textit{Short} and \textit{Quad}-based baselines. \textbf{Row 4:} Competing baselines fail to recover the straight fencing in front of the white car fully, while \textit{L-SVPE} is able to reconstruct the fencing covering the entire car.}
    \label{fig:baselines_img}
\end{figure*}

In Figure \ref{fig:baselines_img}, we present a few qualitative examples of reconstructions with the top-performing baselines. 
To produce reconstructed RGB images, we individually reconstruct each channel with our baselines, which are trained on grayscale images. L-SVPE generalizes well across different color channels, while methods such as Short and Quad Bilinear can introduce discolored artifacts. 
We highlight the robust performance of our model against optical blur observed in Row~2 and global blur observed in Row~3. Row~1 and Row~4 demonstrate how L-SVPE is capable of recovering high frequency detail in the wrinkles of the coat and the fencing in front of the white car, respectively.  

\section{Prototype}
We implement a physical prototype of our model using a focal-plane sensor--processor. Specifically, we use the SCAMP-5~\cite{carey2013100}, a $256 \times 256$ pixel array in which each pixel contains a programmable processing element (PE). Each PE contains a small number of analog and single-bit digital memories that can be set using dedicated instructions. We program the learned coded exposure pixel-wise into the analog memories using a micro-controller.

\begin{figure*}[tp!]
    \begin{center}
    \includegraphics[width=0.75\textwidth]{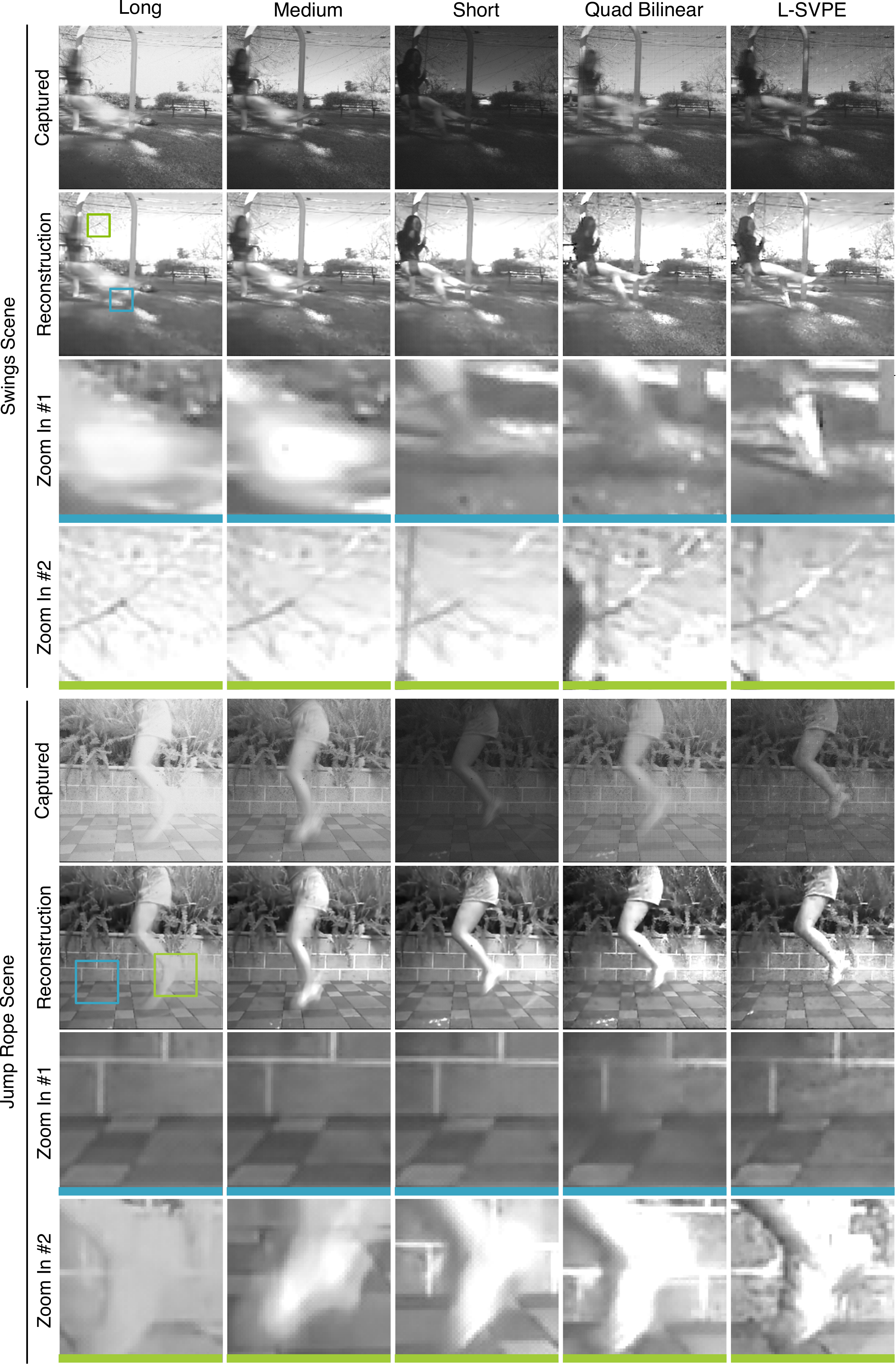}
    \end{center}
       \caption{Reconstructions of images captured using a physical focal-plane sensor--processor prototype (SCAMP-5). We compare reconstructions of two scenes (\textit{Swings} and \textit{Jump Rope}) from global exposures (\textit{Long}, \textit{Medium}, and \textit{Short}) models and the best performing spatially varying exposures (\textit{Quad} with Bilinear interpolation and L-SVPE). 
       \textbf{Rows 1} and \textbf{5}: Captured images. The Short capture is noisier than its longer exposure counterparts. 
       \textbf{Rows 2} and \textbf{6}: Reconstruction from the networks. 
       \textbf{Rows 3} and \textbf{4}: Zoom ins of the reconstructed \textit{Swings} scene. 
       \textit{L-SVPE} can successfully recover the lower shoe in addition to maintaining the high frequency detail of the static tree. 
       \textbf{Rows 7} and \textbf{8}: Zoom ins of the reconstructed \textit{Jump Rope}. 
       \textit{L-SVPE} preserves the sharp edges of the ground tiles and reconstructs the details of the shoe.}
    \label{fig:prototype}
\end{figure*}

Figure \ref{fig:prototype} shows reconstructions from trained global exposure models (Long, Medium, and Short) and the two top-performing spatially varying exposures (Quad with Bilinear interpolation and L-SVPE) on two captured scenes, \textit{Swings} and \textit{Jump Rope}. 
Although Long and Medium reconstructions preserve the static background details, there are few improvements in motion deblurring for both scenes. The Short model tends to blur high frequency details such as the grass in \textit{Swings} and the shoe detail in \textit{Jump Rope} in an attempt to denoise the image. 
Quad Bilinear introduces blurring in areas with fine detail, losing details in the overhead wires in \textit{Swings} and the white edges of the tile in \textit{Jump Rope}. 
Only L-SVPE can successfully recover the fine details of each scene in addition to deblurring. 
More details on how the captures were processed and videos of these scenes can be found in the supplemental material.

\section{Discussion}
We present a novel method for motion deblurring using learned coded pixel exposures. 
We demonstrate that our joint hardware-software approach is better than deep learning for comparable reconstruction networks.
L-SVPE is able to address dynamic motion blurs and varying levels of noise across many different scenes. 
We demonstrate that its performance translates to a physical prototype, in which we show that L-SVPE can deblur while preserving high frequency details in real-world scenes.%

\textbf{Limitations}
Our programmable sensor operates in grayscale, without any color filter on top of the sensor. 
Thus, we design our method around capturing grayscale scenes and process RGB channels individually. 
Additionally, many motion blur datasets do not contain many samples that would allow robust training against over- and under-exposed scenes. 
To our knowledge, no dataset with sufficient motion blur captured at a high frame rate exists. 
Therefore, due to data limitations, we do not test against these lighting scenarios.

We could alternatively use an arbitrarily short exposure to mitigate motion deblurring and focus solely on denoising. 
However, shorter exposures suffer from quantization artifacts on the sensor and require extensive denoising algorithms~\cite{chen2018learning,wei2020physics,tassano2020fastdvdnet,jiang2019learning}. 
Thus, our method is a more robust solution to different lighting scenarios.

\textbf{Future Work}
In future work, we would like to address the aforementioned limitations. 
Such work would include incorporating HDR scenes so that we can train for over- and under-exposed scenes. 
We also do not explicitly focus on optical blur or defocus blurs, and thus an improved E2E model could include modeling the optics for improved robustness against different types of blurs. 
It would be additionally useful to expand E2E methods like ours to programmable sensors that can capture different color channels as well. 
Different colors captured at different exposures such as that of Jiang et al.~\cite{jiang2021hdr} can provide helpful cues in reconstruction.

\textbf{Conclusion}
These efforts add to the growing foundation for emerging programmable sensors in computational imaging. 
As these sensors become more widespread in their use, we may begin to reframe our thinking of how to address ill-posed computer vision tasks, from object classification to HDR imaging. 
This work serves as a step in that direction.

\ifpeerreview \else
\section*{Acknowledgments}
This project was in part supported by NSF Award 1839974, an NSF Graduate Research Fellowship (DGE-1656518), a PECASE by the ARL, and SK Hynix. We thank Piotr Dudek for providing the SCAMP-5 sensor and related discussions. The authors would also like to thank Mark Nishimura, Alexander Bergman, and Nitish Padmanaban for thoughtful discussions. 
\fi

\bibliographystyle{IEEEtran}
\bibliography{references}

\newpage

\section*{Supplementary Material}
\renewcommand{\thesection}{\Alph{section}}
\setcounter{section}{0}

\counterwithin{figure}{section}
\counterwithin{table}{section}
\renewcommand\thefigure{\thesection\arabic{figure}}
\renewcommand\thetable{\thesection\arabic{table}}
\section{Network Details}
\subsection{Simulating noise}
Given that we are simulating capture on a sensor, we take care to add shot noise, a Poisson process that is signal-dependent, and read noise, a Gaussian-approximated process independent of the signal. Following Mildenhall et al.~\cite{mildenhall2018burst}, we model both noise sources as a Gaussian distribution:

\begin{equation}
\label{eqn:noise}
    \mathbf{y}_p \sim \mathcal{N}(\mathbf{x}_p, \sigma_r^2 + \sigma_s \mathbf{x}_p),
\end{equation}
where $\mathbf{y}_p$ represents the noisy measurement of the true intensity $\mathbf{x}$ in pixel location $p$. We vary the noise parameters uniformly across images to simulate changes in sensor gain, and use the gain values as presented in~\cite{mildenhall2018burst} ($\sigma_r$ = [1e-3, 3e-2], $\sigma_s$ = [1e-4, 1e-2]). An accurate mixed noise model~\cite{foi2008practical} allows us to more closely simulate sensor capture over models that use overly simplified additive Gaussian noise. We then clip all measurements to be within $[0, 1]$ to simulate sensor quantization. 

We believe the interpolation step is, in fact, a bottleneck to achieving the model's full reconstruction potential. Just as one uses color-specific gradients in classic debayering of RGB images, one can imagine that a similar exposure-specific interpolation can be beneficial here. We note an exposure-specific interpolation algorithm may work better than our naive Bilinear and Scatter interpolation. However, the design of such an algorithm is beyond the scope of this paper.

\subsection{Decoding U-Net}
The 2D U-Net used takes as input $C$-channel images of resolution $512 \times 512$ to produce a single channel output of the same resolution as its input. The architecture used consists of 6 downsampling blocks and 6 upsampling modules. Each downsampling block consists of a single \texttt{Conv} operation with $3 \times 3$ kernels, followed by a \texttt{ReLU}. The convolution uses a striding of size 1 and padding of size 1. The upsampling module uses \texttt{ConvTranspose} with kernel size $4 \times 4$, stride size 2, and padding size 1. This is followed by 2 \texttt{Conv} operations with the same initialization as the those of the downsampling operation. We then apply a final \texttt{Conv} operation with kernel size $1 \times 1$.
We found that \texttt{BatchNorm} decreased overall performance.

\subsection{Using alternative decoders}
 We sought decoder alternatives that were comparably less demanding of memory and computation than the U-Net. We attempted training a model with the first two stages of MPRNet~\cite{zamir2021multi}, as the entire network did not fit on a single 24 GB GPU. We found that the MPRNet performed comparably well as the U-Net, but required five times as much memory and thus was not our network of choice.
 
As part of our ablation study, we use DnCNN~\cite{zhang2017beyond}. The head consists of a \texttt{Conv} operation with a $3 \times 3$ kernel, with stride size 1 and padding size 1, followed by a \texttt{ReLU}. The body consists of 15 ``blocks,'' each consisting of a \texttt{Conv} with the same parameters as the head, followed by a \texttt{BatchNorm} with momentum 0.9 and a \texttt{ReLU}. The tail consists of a single \texttt{Conv} operation. The network is structured to predict the residual of the input, and we subtract the predicted residual from the 4th channel. This channel corresponds to the shortest channel in the Quad exposure. 

\section{Training Details}
\subsection{Learning a discretized exposure}
Our model discretizes the exposure time per pixel so that it can be programmed in-pixel to our focal-plane sensor--processor, the SCAMP-5~\cite{carey2013100}.
Each pixel is limited to 7 analog memories and 13 single-bit digital memories. 
We use 6 of the digital memories to create $2^6 = 64$ possible time-slots, where 64 out of 64 ``on'' encoded bits would be equivalent to our Long exposure (32 for Medium and 8 for Short). 
We segment the 64 time-slots into 8 to get 8 learnable options for exposure times. 
Here, 8 out of 8 is Long, 4 out of 8 is Medium, and 1 out of 8 would be a Short exposure length.
Learning an exposure length with options 1 through 8 poses a challenge for neural networks, as we are interested in regressing an integer value which is non-differentiable. For this, we use a straight-through estimator (STE)~\cite{bengio2013estimating,hinton2012neural} that allows us to convert our learned exposure values to integer values in the forward function and pass the gradients to these exposure values in the backward pass. Using a STE also allows us to initialize our learned exposure value, contrary to using methods like Gumbel-Softmax which injects Gumbel noise at every iteration.

 \subsection{Training}
 We trained using batch sizes of 2 due to memory constraints on GPUs available for our experiments. We also use \texttt{ReduceLROnPlateau} as a learning rate scheduler with a patience of $20$ and factor of $0.8$ for a gradual decrease in learning. To speed up training, we initialize L-SVPE with the Quad exposure arrangement. 
 
 \subsection{Dataset}
 We segment the videos in the NfS dataset~\cite{kiani2017need} into segments each containing eight frames for two reasons: (1) the average of eight 240-fps frames makes for a single 30 fps which is common in many consumer cameras and (2) the SCAMP-5~\cite{carey2013100}, is limited to 13 digital memories (e.g. $2^13$ time points where the shutter can be ``on'' or ``off''). We choose to use 6 digital memories to get 64 time slots which can be divided into eight frames easily.
 
 We found that the first frame as ground truth improved reconstruction more than using the middle or last frame. We argue this is due to the fact that denoising has been observed to be simpler than blind deblurring~\cite{zhang2010denoising}. When we have a mixture of short and longer exposures, the network can focus on denoising a short exposure to match the ground truth and using static pixels from the longer exposures to supplement the appearance of the short exposure.
 
 We also tested our method on the GoPro Motion Blur dataset~\cite{nah2017deep}. However, we found that given that the dataset was created by deliberately creating motion via rotation of the camera axis, many of the training videos only had global motion. In cases of extensive global motion, the long exposure becomes less helpful in reconstruction. We were interested in have a more generalized method towards global, local, and no motion and thus found the NFS dataset more optimal for our purposes.

\section{Experiment Details}
\subsection{Baselines}
We now discuss in detail our reasoning behind the design of our baselines. Note that \textbf{B} represents Bilinear interpolation and \textbf{S} represents Scatter interpolation. 

\subsubsection{Global Exposures}
\begin{itemize}
    \item \textbf{Burst Average}: We perform a per-pixel averaging of all the frames in the scene, equivalently to a Long exposure or a 30 fps capture, which performs temporal denoising but not spatial denoising. This baseline is subject to misalginment if there is a dynamic motion. The averaged image is then used to compute metrics, skipping network reconstruction completely. This baseline serves as the result of not using any deep learning.
    \item \textbf{Short}: We use the first frame of the video segment input as the short exposure (240 fps). In this case, the decoding network simply must scale and denoise the image, as the ground truth is also the first frame. We found that denoising in general can be easier than deblurring. However, low light instances will prompt the network to blur to denoise, losing critical high frequency detail.
    \item \textbf{Medium}: We average the first four frames to simulate a medium exposure (120 fps), an intermediate choice of possible lengths.
    \item \textbf{Long}: Much like the Burst Average, we perform a per-pixel average of all 8 frames (30 fps), but we then use a decoder for refinement and spatial denoising.
    \item \textbf{Full}: This baseline is a stack of the Short, Medium, and Long exposure at full resolution all captured with the same start time and same viewpoint, which is physically impossible to capture but easy to simulate.
\end{itemize}

Given an 8-frame long scene input, we sought to test a number of global exposures that could be representative of not only ones you may capture on a camera but also could be feasibly implemented on the SCAMP-5 sensor. 
We believe testing each individual Short, Medium, and Long exposure lengths would give us a sufficient grasp of how the network performs on different exposure lengths, and found that, as expected, they provided a linear relationship in performance. The longer the exposure, the more susceptible the exposure was to misalignment, and the worse the reconstruction became. 
The Full exposure is physically infeasible to capture at full resolution, but we believe that the combination of the Short, Medium, and Long exposures at full resolution would serve as useful theoretical upper bound. This baseline demonstrates the utility of information from additional exposures. 
Based on Table \ref{table:baselines}, we see that the Full is able to outperform any individual global exposure by themselves.

\subsubsection{Coded Exposures}
\begin{itemize}
    \item \textbf{Uniformly Random (S)}: We uniformly sample each pixel between exposure lengths 1 to 8. We chose a uniformly random array to demonstrate the worst-case scenario of using multiple exposures. The spatial variability of the exposure would make it challenging for the decoding network to learn a spatially invariant kernel to apply across the measurement. We use PyTorch's \texttt{torch.randint} function to randomly select integers from 1 to 8 to populate a empty $512 \times 512$ array. We do not use length 0 because a pixel that is ``off'' does not provide any additional information for reconstruction. 
    \item \textbf{Poisson Random (S)}: This baseline serves as a non-ideal scenario similar to the Uniform Random exposure due to the spatial variability. However, the Poisson sampling guarantees that in every $3 \times 3$ grid of the $512 \times 512$ array, there is an equal likelihood of finding a pixel of exposure lengths 1 through 8. We believe that having a more equal distribution of the different lengths would be useful for interpolating exposures that were more reflective of the entire scene. We use a multi-class Poisson disk sampling algorithm~\cite{wei2010multi} to populate an empty $512 \times 512$. Given the dense sampling of our array, there will be remaining unfilled pixels that do not meet the radial constraint of the multi-class Poisson sampling. These remaining pixels (we usually had $\sim$20 unfilled pixels out of a total of $512 \times 512$ pixels) are manually filled in to maintain an even distribution of exposure lengths.
    \item \textbf{Nonad (nine-tuple) (B, S)}: This baseline serves as a regular arrangement of the full spectrum of exposure lengths. The learned decoding kernel of the decoding network could then be applied spatially invariantly. We use PyTorch's \texttt{torch.randperm} to randomly arrange an array of pixel exposures, each pixel a unique length from 1 through 8, with an additional exposure length 1 to make a total of 9 pixels in the $3 \times 3$. We then tile this fixed arrangement. We add a short exposure pixel as the last pixel due to the ease of denoising the short exposure over deblurring.
    \item \textbf{Quad (B, S)}: This baseline is designed to resemble the work closest to ours from Jiang et al.~\cite{jiang2021hdr}. Notably, they use a custom decoder network for their reconstruction. However, we were interested in using a decoder network that could be applied fairly well to other baselines with less parameters so we opt for the U-Net. Following Jiang et al.~\cite{jiang2021hdr}, we use a fixed tile coded arrangement of LMMS (Long-Medium-Medium-Short). We then tile this fixed arrangement across the $512 \times 512$ sensor.
\end{itemize}

\begin{figure*}[tp!]
    \begin{center}
    \includegraphics[width=\textwidth]{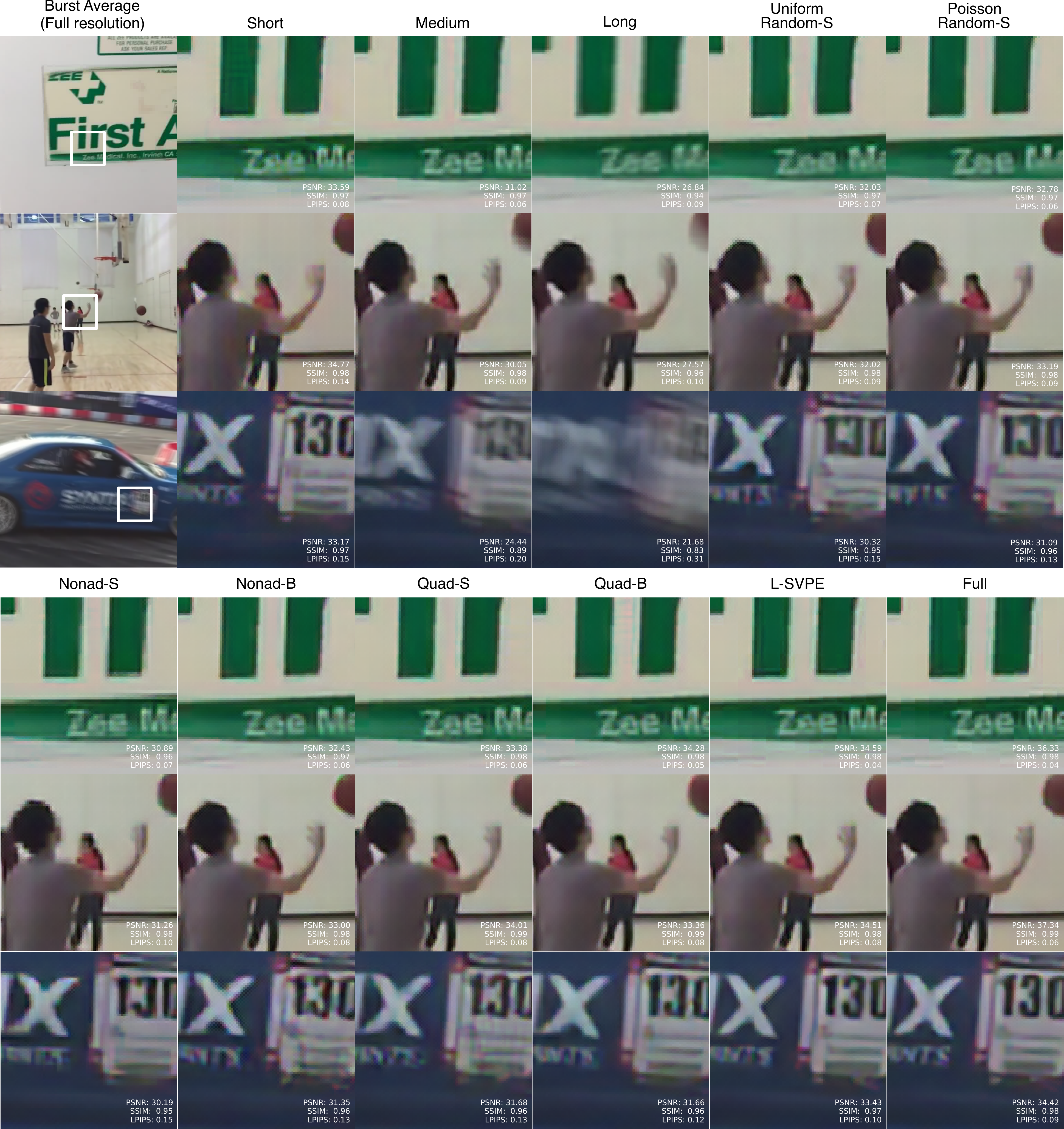}
    \end{center}
    \caption{Qualitative comparison of baselines. Each row of individual images (carried over to create two larger rows) shows an example frame of the average of all the video segment frames (\textit{Burst Average}) at full resolution and a $100 \times 100$ crop of a region for all baselines.
    \textbf{Row 1:} An example of a little to no motion scene. The lettering is most clearly recovered and free of color artifacts with \textit{L-SVPE} and \textit{Full}. 
    \textbf{Row 2:} An example of a scene with local motion. The reconstruction of the man's hand and the deblurring of the basketball is best observed in spatially varying coded exposures. \textit{L-SVPE} is also relatively free of color artifacts.
    \textbf{Row 3:} An example of a scene with global motion. The ``X'' on the side of the car and the lines on the logo next it are clearly recovered by \textit{L-SVPE} while methods like \textit{Short} and \textit{Quad} introduce more blurring artifacts.}
    \label{fig:grid_compare}
\end{figure*}

We compare the Scatter and Bilinear interpolation of these coded exposures where applicable to demonstrate the utility of the interpolation step in terms of perceptual quality. 
We provide the full numerical results from the main paper in Table \ref{table:baselines}, which uses our default parameters including our custom perceptual loss. 
We also provide qualitative comparisons of all the baselines in Figure \ref{fig:grid_compare}. 
To generate RGB images, we run each channel through the decoding network trained on grayscale images. We find that L-SVPE generalizes across different color channels best, after Full, in addition to reconstructing high frequency detail and deblurring.

\setlength{\tabcolsep}{2pt}
\begin{table}
\begin{center}
\caption{Quantitative results of all baselines. \textit{Full}, our theoretical upper bound performs the best on all three metrics, and \textit{L-SVPE} performs the best out of all other baselines on all metrics.}
\label{table:baselines}
\begin{tabular}{l|c|ccc}
\hline\noalign{\smallskip}
Model & Interpolation & PSNR$\uparrow$ & SSIM$\uparrow$ & LPIPS$\downarrow$\\
\noalign{\smallskip}
\hline
\noalign{\smallskip}
Burst Average & ---  & 25.305 & 0.827 & 0.462     \\
Short & ---  & 33.618 & 0.949 & 0.158\\
Medium & ---  & 28.669 & 0.919 & 0.165 \\
Long & ---  & 25.666 & 0.879 & 0.215 \\
Uniform Random & Scatter & 31.607 & 0.946 & 0.139 \\
Poisson Random & Scatter & 32.098 & 0.950 & 0.136\\
Nonad & Scatter & 31.614 & 0.943 & 0.142\\
Nonad & Bilinear & 32.731 & 0.954 & 0.125 \\
Quad~\cite{jiang2021hdr} & Scatter & 33.272 & 0.957 & 0.122 \\
Quad & Bilinear & 33.632 & 0.962 & \underline{0.111} \\
L-SVPE & Scatter & \underline{34.383} & \underline{0.963} & \underline{0.111} \\
Full & ---  & \textbf{36.375} & \textbf{0.976} & \textbf{0.085}\\
\hline
\end{tabular}

\end{center}
\end{table}
\setlength{\tabcolsep}{1.4pt}

\section{Prototype}
The physical prototype, a SCAMP-5 focal-plane sensor--processor~\cite{carey2013100}, we used has a non-linear camera response function (CRF). We first calibrate the CRF using a set of exposures and the CRF function from OpenCV. We use the CRF to linearize our captures before processing. 

The SCAMP-5 has a sensor resolution of $256 \times 256$. However, we trained all of our models at $512 \times 512$ resolution. Thus, to reconstruct the images captured from the SCAMP-5, we pad these images with zeros before in putting them into our network. We then apply an sRGB curve ($\gamma = 2.0)$ to the reconstruction for display. 

The supplemental videos of the scene show the frame-by-frame reconstruction of the \textit{Swings} and \textit{Jump Rope} scene. Although the capture itself is reflective of the fps designated for each model, the recording is saved at 60 fps due to the design of the camera software. We also slow down the video to 5 fps to allow for viewing detail. 

\end{document}